\documentclass[12pt]{iopart}
\usepackage{iopams}  
\usepackage{graphicx}

\newcommand{\ets}{\boldsymbol{\eta}}
\newcommand{\eqref}{\ref}

\begin{document}

\title[Mechanics of nematic membranes]{Mechanics of nematic membranes:
 Euler-Lagrange equations, Noether charges, stress, torque and boundary conditions of the surface Frank's nematic field}
\author{J  A Santiago$^{1, 2}$ and F Monroy$^{2, 3}$}
\address{$^1$ Departamento de Matem\'aticas Aplicadas y Sistemas, Universidad Aut\'onoma Metropolitana Cuaijimalpa, 
Vasco de Quiroga 4871, 05348 Ciudad de M\'exico, MEXICO}
\address{$^2$ Departamento de Qu\'imica F\'isica, 
Universidad Complutense de Madrid, Ciudad Universitaria s/n, 28040, Madrid, SPAIN}
\address{$^3$ Institute for Biomedical Research Hospital Doce de Octubre (imas12)\\
Av. Andaluc\'ia s/n 28041, Madrid, SPAIN}
\eads{\mailto{jsantiago@correo.cua.uam.mx},
\mailto{monroy@ucm.es} }

\vspace{10pt}

\begin{abstract}
The mechanics of a  flexible membrane decorated with a nematic liquid-crystal texture is considered in a variational framework. 
The variations  on the splay, twist and the bend  energy of the nematics are obtained from the local deformations 
leading to changes in  the shape membrane.   The Euler-Lagrange derivatives and the  Noether charges 
are identified from the variational equations.  The nematic stress tensor is obtained as a consequence of translational invariance. Likewise, the rotational invariance implies the torque nematic tensor. The corresponding boundary conditions are obtained for free edges in the open-membrane configuration. These results constitute the basis of a generalized theory of elasticity for anisotropic nematic membranes. Some relevant consequences of the presence of nematic ordering are visualized at revolution surfaces with  axial symmetry.
\end{abstract}

%
\noindent{\it keywords\/}: Frank's nematic energy, Canham-Helfrich energy, nematic ordering, elastic membrane, nematic stress tensor, surface vector field.
%
%
%
%

\section{Introduction}
\begin{figure}[th]
\centering 
\includegraphics[width=3.4in]{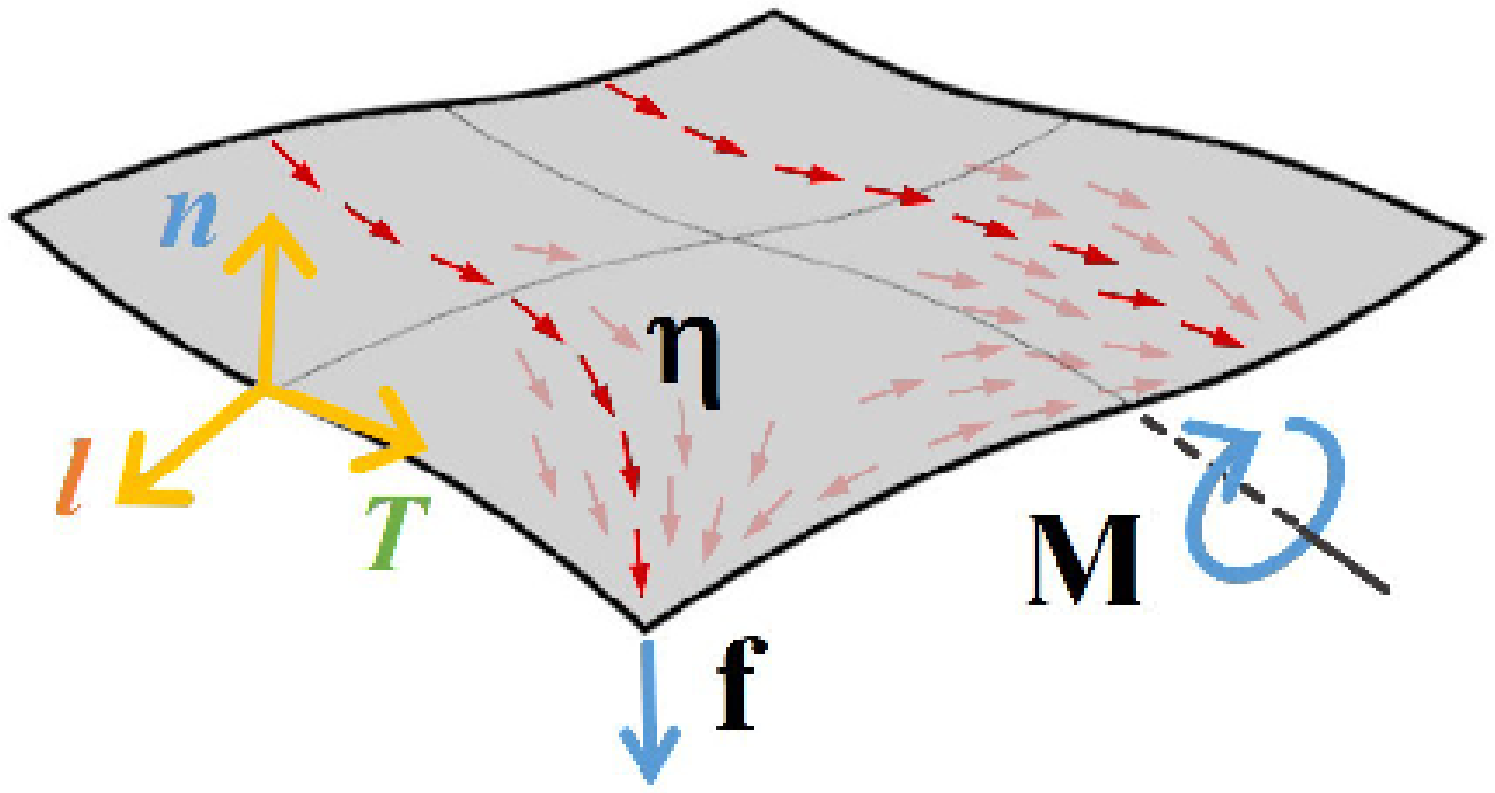}
\vspace{-2mm}
\caption{Flexible membrane coated with a surface nematic texture
represented by a molecular director $\ets$ of unitary length ($\ets\cdot\ets$ = 1, equivalent to the homogeneity
property of the liquid crystal), which is forced to remain tangent to the surface ($\ets\cdot {\bf n} = 0$, 
endowing strict two-dimensional character to the surface nematics). ${\bf M}$ represents the intrinsic torque tensor 
and ${\bf f}$ the stress tensor.
The Darboux frame on the edge curve is shown, where ${\bf l}= {\bf T}\times {\bf n}$.  }
\vspace{-2mm}
\label{FF1}
\end{figure}

The concept of mechanical equilibrium is a cornerstone in understanding the physics of elastic membranes \cite{lipowski, nelsonbook}. The Canham-Helfrich (CH) theory of the curvature-elasticity of isotropically fluid membranes recapitulates the relevant degrees of freedom involved in the elastic energy due to curvature \cite{canham, helfrich}. The variational minimization of the CH-functional has been successfully exploited to describe  equilibrium shapes 
of fluid membranes upon given constraints \cite{seifert}. The particular case of an isotropic fluid membrane embedded in a nematic solvent has been theoretically addressed from the point of view of the membrane thermal fluctuations, which appear somewhat restrained by the more ordered 3D medium \cite{terenjev}.  More recently, the CH-theory has been also used to study the distribution of membrane stresses and the induced torque calculated along the fluid membrane \cite{guvenbook, stress, deserno, fournier}. In this mechanical context, a theory of elastic membranes coated with a nematic texture has not been formulated so far \cite{napoli2010, napoli2018}.  To cartoon the problem, Figure \ref{FF1} 
shows  the anisotropic organization of the elongated molecules that constitute the membrane nematics. The related property of tangential orientation (in-plane nematics) is described by a unitary vector field, the so-called nematic Frank director $\ets$, which specifies the direction of the 
nematic order parameter \cite{lubensky-book}. We are actually considering a zero-temperature system of anisotropic objects ordered in a two-dimensional geometry that potentially orientates the objects into preferential directions along the field of membrane forces as represented
by the stress $\bf f$ and torque $\bf M$. Although thermal fluctuations are expected to destroy long-range order in two dimensions \cite{peierls}, the nematic membranes here considered are spatially finite to preserve orientational correlations  and often subjected to constraining boundary conditions thus enabling for topological order even at high-temperature \cite{thouless1, thouless2}. 
In infinite 2D-systems, Peierls argued that no long-range order exists because thermal phonons move atoms from their equilibrium positions with a mean square displacement that increases logarithmically with the size of the system \cite{peierls}. The absence of long range order of this simple form was shown rigorously by Mermin \cite{mermin}. This situation is radically different in finite systems at zero-temperature with short-range orientational ordering which is highly enforced by the existence of boundary conditions that constrain the internal structure of the nematic vector field (see Fig.  \ref{FF1}).

Furthermore, mutually-interacting  topological defects could eventually appear in closed  membranes, and then self-organize because of the necessary geometrical congruence of the intrinsic director field with the membrane geometry of the membrane and the internal structure of the mechanical force field \cite{mac,bowik, kamien, santiagog, selinger}. Because the presence of orientational interactions could impose a preferred orientation, we assume $\ets$ as an arrowed vector pointing in the specific direction of the field of membrane forces (see Fig. \ref{FF1}). 
In equilibrium, the static configuration  of the zero-temperature nematic field  should be determined by the spatial distribution of the membrane stresses, through the splaying, twisting and bending of the nematic director \cite{gennes},  in a tight interplay with the underlying CH-elastic forces. All these forces are intrinsically coupled each other in connection with the membrane geometry, which encodes how the mechanical information contained on the director field should distribute along the nematic membrane.

In a previous paper, from a geometric standpoint we approached the unresolved problem of a flexible membrane with a tangentially embedded 
nematic field \cite{monroy}. By considering the Frank's energy functional particularized to the two-dimensional case, the surface distribution of membrane stresses and torques were analyzed in terms of the intrinsic and extrinsic counterparts of the membrane curvature. Then,  the different distortion rigidities of the molecular director were adapted to the surface geometry to render into the global structure of the membrane nematic field.  Finally, the geometric characteristics of the nematic vector, i.e. surface tangentiality and unitarity, were superposed a posteriori as constraints to the director field \cite{monroy}. The problem was resolved using auxiliary variables that introduce Lagrange multipliers associated with constraints of the surface geometry as embedded into the Euclidean space \cite{guven-auxiliary}. Whereas that method had the advantage to avoid cumbersome calculations of membrane deformations in every one of the terms in the Frank's energy functional, now we propose to address a generalized theory of elasticity that determines the local distribution of the nematic membrane stresses in terms of a generalized strain field that is a common thread to all of them. The resulting theory will be hence completely covariant as affinely connects the vector director field with the embedding surface geometry that represents the curvature field of the membrane. As a motivation, Figure~\ref{FF1} depicts how the ordered configuration of the membrane-embedded nematics should impose 
anisotropic force distributions inherently coupled to the "flexible" geometry of the membrane as determined by its curvature field, similarly to the CH-theory of fluid membranes \cite{guvenbook, stress, deserno}. Looking specifically at the anisotropic stresses due to curvature-driven distortions in the director field, one could observe either a spatial distribution arising from the geometric coupling with the local curvature or, conversely, a director field locally reorganized upon geometric remodeling. The present study explores a genuine mechanical route undergone through of the variational principle implemented on an intrinsically surface-embedded nematic setting constructed {\it a~priori}.  Although this mechanical pathway will have the operational disadvantage to handle many coupled terms before the final result is attained, it enables to enlighten the several variations of the Frank's energy appeared upon deformation, which naturally give rise to the different membrane forces due to internal equilibrium terms and boundary generators of their conserved currents~\cite{santiagog}. This route is quite different to the auxiliary method previously exploited \cite{monroy}, which delivers the equilibrium forces without specific detail of their sources.
Here, we will take advantage of differential geometry to calculate how the surface-adapted Frank energy respond to infinitesimal deformations of the membrane shape. Expanding up to first order deformation, the energy variation delivers the Euler-Lagrange derivative plus boundary terms arising from  the Frank's response of the nematic membrane. In equilibrium, the deformations of the energy are identically the boundary terms, the so-called Noether charges,  which are by themselves invariant under the corresponding symmetry transformation. For every continuous symmetry transformation possible in a flexible membrane, the Noether's theorem declares the existence of conserved Noether charges \cite{guvenbook}. We will identify the conservation of the stress tensor $\bf f$ as the conserved current of an invariant charge that preserves translation invariance at the membrane boundary; similarly, rotational invariance applies for membrane torque {\bf M}.    Intuitively, in a soap film, the tension stress is  the Noether charge that generates the lateral current that conserves the surface area, similarly to the gauge symmetry in electromagnetism, which generates the conserved electric current. In our theory, the Noether charges are among the more important characteristics of the nematic membrane; they  determine its physical state allowing the derivation of the forces that act on the membrane boundaries by using the invariance of the Frank's energy under rigid motions in space, for instance, translations for {\bf f} and rotations for {\bf M}. Because the Frank's energy remains invariant under these symmetries, the corresponding Noether charges become conserved along the membrane.  With respect to the exogenous pathway followed in \cite{monroy} via auxiliary variables, and despite  the radically different way of formulating the current endogenous theory with affine connections between nematic and curvature fields, here we will  reach exactly the same results for the stress and torque tensors, a fact evidencing the formal equivalence between the two theories.
The rest of the paper is organized as follows:
In Section \ref{DND}, we  briefly summarize the fundamentals  needed to calculate the
deformations of the nematic director. 
In Section \ref{FFEE} we outline the general expression for  Euler-Lagrange equations and the Noether charges. 
The specific results for splay, twist and bend are made explicit in Section \ref{SPLAY}, \ref{TWIST} and \ref{BEND}, respectively. 
For the sake of example, the particular case of an axially symmetric vesicle with a nematic director aligned on meridians is worked out
in Section \ref{AXIAL}.
A discussion of the results and an outlook of their practical implications  are given in Section \ref{DIS}.  Finally,  the conclusions are summarized in Section \ref{CON}.
The most important details on the calculation of the deformations are presented in several appendices. 
\section{Deformation of the nematic director}\label{DND}
We describe the membrane as a surface embedded in $R^3$,
using the
embedding functions 
\begin{equation}
{\bf x}= {\bf X}(\xi^a),
\end{equation}
which is parametrized by two parameters $\xi^a$, $a=1, 2$; here ${\bf x}=(x^1, x^2, x^3 )$, the cartesian coordinates
of $R^3$. The infinitesimal distance on the surface
\begin{equation}
ds^2=g_{ab}d\xi^a d\xi^b,
\end{equation}
is defined in terms of the induced metric $g_{ab}={\bf e}_a\cdot {\bf e}_b$, where
${\bf e}_a=\partial_a {\bf X}$, are two tangent vectors to the surface~\cite{docarmo}.
Here, the metric $g_{ab}$ and its inverse $g^{ab}$ are used to raise and lower surface
indices in the geometric objets.

Nematic textures imprinted on the surface can be described by a unit vector field, 
the so-called nematic director $\ets$ \cite{gennes}; in the present context, this is an unitary nematic vector, which is assumed normalized, 
$\ets\cdot \ets=1$ and  forced to lie
tangent to the surface (see Fig. \eqref{FF1}). Consequently,  the membrane-embedded molecular vector that represents the oriented surface nematic can 
be written as  $\ets=\eta^a{\bf e}_a$, where $\eta^a=\ets\cdot {\bf e}^a$ are the
projections on the two tangent vector fields. Equivalently, given the tangential nature of the membrane nematic,
we can establish the condition $\ets \cdot {\bf n}=0$, where $\bf n$ is the unit normal to the surface, this is ${\bf n}= {\bf e}_1\times {\bf e}_2/\sqrt{g}.$

\subsection{Generalized deformation: Strain field.} 
Any infinitesimal deformation of the embedding functions, $\delta{\bf X}$, can be projected along the surface as
a strain field with two components, in-plane deformations $\Phi_a= \delta{\bf  X}\cdot {\bf e}_a$,  and  the normal 
deformations  $\Phi=\delta{\bf X}\cdot {\bf n}$; therefore,  we can write:
\begin{equation}
\delta {\bf X}= \Phi^a{\bf e}_a + \Phi {\bf n}.\label{DEFOR}
\end{equation}
Deformation of the unitary relation $\ets \cdot \ets=1$  gives $\delta \ets \cdot \ets=0$, thus considering that
the nematic director  is tangential to the surface $\ets=\eta^a{\bf e}_a$, we 
find $\delta \ets \cdot {\bf e}_a=0$.  Therefore, 
if we expand  $\delta\ets$ in the local basis $\{ {\bf e}_a, {\bf n} \}$, one gets
\begin{equation}
\delta\ets= (\delta \ets \cdot {\bf e}^a) {\bf e}_a + (\delta \ets \cdot {\bf n}) {\bf n},
\end{equation}
where the first term strictly vanishes, i.e. within the current theory, any variation of the nematic director 
is strictly normal to the surface. Moreover, the deformation of the nematic director 
can be written 
in terms of $\delta {\bf n}$ as
\begin{eqnarray}
\delta\ets &=& (\delta\ets \cdot{\bf n} ) {\bf n}, \nonumber\\
&=&- (\ets\cdot \delta {\bf n} ) {\bf n}, 
\end{eqnarray}
where we have considered the relationship $\delta\ets \cdot {\bf n}= -\ets\cdot \delta{\bf n}$.
Therefore, since 
$\delta {\bf n} = (-\nabla^b \Phi + K_a{}^b\Phi^a ) {\bf e}_ b$ \cite{santiago-defor}, 
we finally find:
\begin{equation}
\delta\ets = - \eta^a (   K_{ab}  \Phi^b - \nabla_a \Phi  ) {\bf n}. \label{WWE}
\end{equation}
In these equations, we recall the symmetric tensor  $K_{ab}={\bf e}_a\cdot \nabla_b {\bf n}$ to be
the extrinsic curvature of the surface, where $\nabla_a$ denotes the covariant derivative, compatible 
with the induced metric~\cite{docarmo}. 
Deformations of the membrane shape imply
deformations of the nematic director according to Eq. (\eqref{WWE});
notice that in this equation, tangential deformations are exclusively 
coupled with the extrinsic curvature. Therefore, surface derivatives of the nematic director can be developed 
within this framework as:
\begin{equation}
\nabla_a \ets= \nabla_a \eta^b {\bf e}_b - \eta^b K_{ab} {\bf n}.
\end{equation}
Let us notice that tangential components are the covariant derivatives of the components $\eta^a$, 
whereas the normal component of this divergence field  couples  with the extrinsic curvature.
Deformation of  this derivative gives rise to
\begin{eqnarray}
 \delta\nabla_b \ets &= & \Big[ \delta (\nabla_b \eta^d ) 
+ \nabla_b\eta^c ( \nabla_c \Phi^d +\Phi K_c{}^d)  
-  \eta^c K_{bc} ( - \nabla^d\Phi  + K_a{}^d \Phi^a    ) \Big] {\bf e}_d\nonumber\\
&+&\Big[ \nabla_b \eta^c (  \nabla_c\Phi - K_{cd}\Phi^d  )  - \delta (\eta^c K_{bc} ) \Big]{\bf n}, \nonumber\\
\label{DDNS}
\end{eqnarray}
where the deformation $\delta(\nabla_a\eta^b)$ has been expanded  in  \ref{SSPP}.
\section{Frank's Nematic energy}\label{FFEE}
Our mechanical theory grounds on the general description of the curvature-elasticity of molecularly 
uniaxial liquid crystals, namely the Frank theory of nematics energy \cite{nematic-frank}. 
The surface nematic  can be modelled by the $2D$-Frank's functional as developed by Napoli and Vergori for 
nematic shells in the limit of zero-thickness \cite{napoli2012}, and in the form as previously implemented in Ref. \cite{monroy}
\begin{eqnarray}
F_{\rm Frank} &=& \frac{\kappa_1}{2} \int dA \,   ( \boldsymbol{\nabla}\cdot \boldsymbol{\eta})^2
+ \frac{\kappa_2}{2} \int dA \,  [ \boldsymbol{\eta}\cdot (\boldsymbol{\nabla} \times \boldsymbol{\eta}) ]^2\nonumber\\ 
& +&\frac{\kappa_3}{2} \int dA\,  [  (\boldsymbol{\eta}\cdot \boldsymbol{\nabla} )  \boldsymbol{\eta}     ]^2,\label{SHELL}
\end{eqnarray}
where $\kappa_1, \kappa_2$ and $\kappa_3$ are the two-dimensional constants for  splay,  twist and bend of 
the nematic membrane. These rigidity constants  retain the mechanical characteristics of the molecular director upon the respective 
distortions occurred in bulk ($\kappa_i = K_i h$), particularized to the surface case of 
vanishing thickness~(i.e. at $ h \to 0 $)~\cite{napoli2012}.  
Here, $dA= \sqrt{g}d\xi^1d\xi^2 $ is the area element on the surface, with $g={\rm det}\, g_{ab}$. 
The $3D$  operators $\boldsymbol\nabla={\bf e}^a\nabla_a $, $\boldsymbol\nabla\times={\bf e}^a \nabla_a \times $
are, respectively,  the surface gradient and the surface curl operators,  which will be   
applied to the nematic director \cite{monroy, napoli}.

\subsection{Variational equations}
The variational effect of the infinitesimal deformations described by Eq. (\eqref{DEFOR}) on the surface Frank's energy, 
can be written  as
\begin{equation}
\delta F_{\rm Frank}= \int dA \, \boldsymbol{\cal E}\cdot \delta {\bf X} + \int dA\, \nabla_a {\cal Q}^a, \label{DDFR}
\end{equation}
where $ \boldsymbol{\cal E} =  {\cal E}_\perp {\bf n} +{\cal E}_a  {\bf e}^a$, 
is the Euler-Lagrange derivative and ${\cal Q}^a$  the Noether charge, which contains both the terms on the two components of the strain field. 
Remarkably,   the tangential components ${\cal E}_a$, do not vanish, because the energy is not invariant under reparametrizations.
Nevertheless, in equilibrium  ${\cal E}_\perp=0={\cal E}_a$.  Then, we take advantage of the Noether theorem to get the conservation laws corresponding to every continuos symmetry. Particularly, we elaborate on the structure of the equilibrium equation upon translation and rotation transformations that preserve the material characteristics of spatial homogeneity and linear flexibility. 

On the one hand, if one considers an infinitesimal translation in a homogeneous membrane $\delta {\bf X}={\bf a}$, 
when looking at the induced variation in membrane energy due to an homogenous deformation,
then, for any area element $dA$, we found:
\begin{equation}
\delta F_{\rm Frank}= {\bf a}\cdot  \int_{\cal M} dA \, [ \boldsymbol{\cal E} - \nabla_a {\bf f}^a].
\end{equation}
Because invariance under  translations, then $\delta F_{\rm Frank}=0$, so we deduce 
\begin{equation}
\boldsymbol{\cal E}= \nabla_a {\bf f}^a,
\end{equation}
where ${\bf f}^a$ is the nematic stress tensor.  Using the divergence theorem we identify the integral 
\begin{equation}
{\bf F}= \int_{\cal M} dA\, \nabla_a {\bf f}^a= \oint_{\cal C} \, ds\,  {\bf f}^a l_a,
\end{equation}
as the force acting on the  edge curve ${\cal C}$, which defines the boundary of the membrane patch ${\cal M}$.

On the other hand, if the membrane is deformed under an infinitesimal rotation 
$\delta{\bf  X}= \boldsymbol{\omega}\times {\bf X}$, we get
\begin{equation}
\delta F_{\rm Frank}= \boldsymbol{\omega} \cdot \int dA\,  [ {\bf X} \times \boldsymbol{\cal E} -\nabla_a{\bf M}^a  ],
\end{equation}
then, rotational invariance implies that
\begin{equation}
{\bf X}\times \boldsymbol{\cal E}  = \nabla_a {\bf M}^a,
\end{equation}
where ${\bf M}^a$ is identified as the nematic torque tensor. In equilibrium, we get the 
covariant conservation of the stress and torque tensors as
$\nabla_a {\bf f}^a=0$ and $\nabla_a {\bf M}^a=0$.
\subsection{Boundary conditions}\label{BBC}
To obtain the boundary conditions, we  consider a membrane patch with  surface tension $\sigma$ and a boundary 
edge characterized by  a line tension $\sigma_b$; the total energy is:
\begin{equation}
F=F_{\rm Frank}  + \sigma \int dA   +\sigma_b \oint ds. 
\end{equation} 
Deformations of the edge curve  can be written as
\begin{eqnarray}
\delta{\bf X}&=& \Phi^a {\bf e}_a + \Phi {\bf n}, \nonumber\\
&=&\Psi_T {\bf T} + \Psi_l {\bf l} + \Phi {\bf n}.\label{STRATAN}
\end{eqnarray}
where $\Psi_T=  \Phi^a T_a$ is the projection of the bulk deformation along the unit tangent  to the edge ${\bf T}$, 
and $\Psi_l = \Phi^a l_a$, the projection along the unit normal $\bf l$ \cite{santiago-edge}. Whether
the boundary  is parametrized by arc length, the corresponding Darboux frame is  
defined as  
\begin{eqnarray}
\dot{\bf T}&=&\kappa_n {\bf n}+ \kappa_g {\bf l},\nonumber\\
\dot{\bf l}&=&- \kappa_g {\bf T}- \tau_g {\bf n},\nonumber\\
\dot{\bf n}&=& -\kappa_n {\bf T}+ \tau_g {\bf l}, \label{DAR}
\end{eqnarray}
where the dot means derivative respect the arc length (see Figure \ref{FF1}) \cite{docarmo}. Here, 
$\kappa_n= \dot{\bf T} \cdot {\bf n} =-K_{ab} T^a T^b$ defines the normal curvature  of the edge curve,
$\kappa_g= \dot{\bf T}\cdot {\bf l} = (\dot T^a + \Gamma^a_{bc} T^b T^c)l_a$ defines its geodesic curvature,
and  $\tau_g=\dot{\bf n}\cdot {\bf l}= K_{ab}T^al^b$ the geodesic torsion.
Finally, we can write 
\begin{eqnarray}
\delta F &=& \int dA ({\cal E}_\perp \Phi + {\cal E}_a\Phi^a ) \nonumber\\
&+& \oint ds\,  [{\cal Q}^a  l_a + (\sigma - \sigma_b \kappa_g)  \Psi_l - \sigma_b \kappa_n \Phi  ], \label{BBCC}
\end{eqnarray}
where ${\cal Q}^a$ is the Noether charge that generates the nematic current as defined in  Eq.~(\eqref{DDFR}); the other two terms represent 
the isotropic   currents due to lateral tensions.
From Eq. (\eqref{BBCC}),  the boundary conditions can be isolated after the Noether charge ${\cal Q}^a$ has 
been properly identified.
In the next sections,  we will obtain explicit expressions for the  Noether charge, stress and torque as well the 
Euler-Lagrange derivatives of the Frank's energy corresponding to each mode of deformation.
\section{Splay energy}\label{SPLAY}
To obtain the deformation of splay energy, we note  that (see \ref{SSPP}): 
\begin{equation}
\delta (\nabla\cdot \ets) = \delta (\nabla_a \eta^a).
\end{equation}
By using Eqs. \eqref{DA} and \eqref{DB}, we get
\begin{eqnarray}
\delta (\nabla_a \eta^a)
&=& \delta_\perp (\nabla_a \eta^a)  + \delta_\parallel (\nabla_a \eta^a), \nonumber\\
&=& -\Phi K_c{}^a \nabla_ a \eta^c + ( K\eta^a - K_c{}^a \eta^c  )\nabla_a \Phi\nonumber\\
&-& \nabla_c\eta^a \nabla_a \Phi^c  -  {\cal R}_G \eta^c\Phi_c. \label{DSS}
\end{eqnarray}
The tangential deformation only includes geometric elements that are intrinsic to the membrane.
Therefore, we can write the deformation of the splay energy as
\begin{eqnarray}
\delta F_S= \frac{\kappa_1}{2}\Big[ \int  (\delta dA) (\nabla_a\eta^a )^2  + 
\int dA\, \delta ( \nabla_a\eta^a)^2 \Big].\label{DER}
\end{eqnarray}
In the first term, the deformation of the area term holds
\begin{equation}
\delta dA = dA \, ( \nabla_a \Phi^a + K\Phi ),
\end{equation}
thus, Eq. (\eqref{DSS}) allows us to obtain
\begin{equation}
\delta F_S= \int dA\, ( {\cal E}_a^S \Phi^a + {\cal E}_\perp^S \Phi ) + \int dA \nabla_a {\cal  Q}^a_S,
\end{equation}
where the expression
\begin{eqnarray}
{\cal E}_\perp^S&=&  \kappa_1\Big[ \frac{1}{2} \nabla_a\eta^a ( K\nabla_a\eta^a  -  K_c{}^b\nabla _b \eta^c)- \nabla_a [  \nabla_b\eta^b (  K\eta^a - K^a{}_c \eta^c ) ] \Big],
\end{eqnarray}
is the {\it normal} Euler-Lagrange derivative, and correspondingly
\begin{eqnarray}
{\cal E}_a^S&=& -\kappa_1 \Big[  \frac{1}{2} \nabla_a ( \nabla_b\eta^b )^2 + (\nabla_b\eta^b) {\cal R}_G\,  \eta_a 
-  \nabla_b ( \nabla_c\eta^c\nabla_a \eta^b )\Big],
\end{eqnarray} 
are the {\it tangential} Euler-Lagrange derivatives. 
Since the divergence of the nematic director does not depend the way the membrane 
is embedded in $R^3$,  ${\cal E}_a^S$ contains only the intrinsic geometric  
information. 
Finally, we can  identify the Noether charge ${\cal Q}^a_S$ as:
\begin{eqnarray}
{\cal Q}^a_S &=&\kappa_1\Big[ \frac{1}{2} (\nabla_b\eta^b)^2 \Phi^a -  \nabla_b\eta^b (\nabla_c \eta^a) \Phi^c 
+\nabla_b\eta^b ( K\eta^a - K^a{}_c \eta^c )\Phi \Big] .
\end{eqnarray}
Here, the first two terms come from the tangential deformations whereas the last one stems on the normal deformation. The pure tangential components of this charge are  proportional to the local density of splay energy but  are curvature-independent, differently to the normal component, which requires a finite curvature coupled to the divergence of the nematic field.
\subsection{Stress and Torque tensor}
The  splay stress tensor can be obtained as a result to apply  an infinitesimal translation  $\delta{\bf X}={\bf a}$ to the splay charge
so that the tangential deformations are $\Phi^a= {\bf a}\cdot {\bf e}^a$,
and the normal deformation, $\Phi= {\bf a}\cdot {\bf n} $.
Accordingly, the  splay stress tensor reads
\begin{eqnarray}
{\bf f}^a_S=\kappa_1 \nabla_d \eta^d \Big[  \left( \nabla^c \eta^a  -
 \frac{g^{ac}}{2}\nabla_d\eta^d \right) {\bf e}_c
-  ( K    \eta^a -  K^a{}_c \eta^c) {\bf n} \Big].
\end{eqnarray}
Notice that the tangential components  of the stress tensor $f^{ab}$
do satisfy the condition $g_{ab}f^{ab}=0$, which is as a consequence of the invariance of the energy under
deformations $\delta {\bf X}= \lambda {\bf X}$ \cite{guvenbook}.
As a matter of fact, if the condition $\nabla_a\eta^a=0$ holds for a given nematic texture, then, 
it does not induce splay stress, which represents a trivial  solution of the Euler-Lagrange equations.\\
To deduce the torque tensor, we now apply an infinitesimal rotation $\delta {\bf X}= \boldsymbol{\omega} \times {\bf X}$, where
$\boldsymbol{\omega}$ is a constant vector. This drives the tangential deformations 
$\Phi^a= \boldsymbol{\omega}\cdot ({\bf X} \times {\bf e}^a)$, 
while the normal deformation $\Phi=\boldsymbol{\omega} \cdot ({\bf X}\times {\bf n} )$.
Finally, the splay torque tensor is  identified as:
\begin{equation}
{\bf M}^a_S= {\bf X}\times {\bf f}^a_S.
\end{equation}

\subsection{Boundary conditions}
As  outlined in Eq. (\eqref{BBCC}), we get the boundary conditions as  the projection of the 
Noether charge on the membrane edge, this is ${\cal Q}^a l_a$. For the splay case we have
\begin{eqnarray}
{\cal Q}^a_S l_a &=&\kappa_1 \left[\frac{1}{2} (\nabla_b\eta^b)^2  - l_a \nabla_b\eta^b l^c (\nabla_c \eta^a) \right]\Psi_l 
-\kappa_1 l_a \nabla_b\eta^b T^c (\nabla_c \eta^a) \Psi_T  \nonumber\\
&+&\kappa_1 \nabla_b\eta^b ( K l_a\eta^a -  K_{ac} l^a\eta^c )\Phi,
\end{eqnarray}
and  using  Eq. (\eqref{BBCC}),  the boundary conditions are identified as
\begin{eqnarray}
 \sigma_b \kappa_g - \sigma &=& \kappa_1\left[  \frac{1}{2} (\nabla_b\eta^b)^2 -
 l_a \nabla_b\eta^b l^c \nabla_c\eta^a\right] , \nonumber\\
\sigma_b \kappa_n &=&  \kappa_1 \nabla_b\eta^b [ K l_a\eta^a -  K_{ac}l^a \eta^c ],\nonumber\\
0&=& \kappa_1 l_a \nabla_b\eta^b  [ T^c\nabla_c \eta^a ]. \label{BBCD}
\end{eqnarray}
We realize that even after the Canham-Helfrich energy has been taken into account (see \ref{HEL}), the last
condition in Eq. (\eqref{BBCD}) remains still valid, so  we have $\dot\eta^a=0$ along the boundary.
In general, if  the divergence of the nematic director,   vanishes along the boundary edge, 
$\nabla_a\eta^a=0$, then the geodesic curvature of the boundary curve 
is a constant $\kappa_g= \sigma/ \sigma_b$, with zero normal curvature, $ \kappa_n=0$.

\section{Twist energy}\label{TWIST}
Now, let us proceed with the twist  energy. Under the definition of a torsional nematic curvature as
$ K_\tau = \ets\cdot \nabla\times\ets =g^{ab} \boldsymbol{\eta}\cdot ( {\bf e}_a \times \nabla_b \boldsymbol{\eta} ),$
the twist energy can be written as
\begin{equation}
F_W = \frac{\kappa_2}{2}\int dA \, K_\tau^2, \nonumber\\
\end{equation}
and its  deformation is given by 
\begin{equation}
\delta F_W= \frac{\kappa_2}{2}\int \delta (dA ) K_\tau^2  + \kappa_2 \int dA\,  K_\tau \delta  K_\tau, 
\end{equation}
Using the results obtained in  \ref{TTSS} and the definition $\ets_\perp= \ets\times {\bf n}$, 
we can write
\begin{equation}
\delta F_W =  \int dA ({\cal E}_\perp^W  \Phi + {\cal E}_a^W \Phi^a )
+ \int dA  \nabla_a {\cal Q}^a_W,
\end{equation}
where we identify the normal Euler-Lagrange derivative as
\begin{eqnarray}
{\cal E}_\perp^W 
&=& - \kappa_2\Big[ \frac{1}{2} K_\tau^2 K 
-   \nabla_a ( K_\tau \eta^d \nabla^a \eta_{\perp d} -2K_\tau \eta^a \nabla_b \eta^b_\perp -\eta^a\eta_\perp^c \nabla_c K_\tau       ) \Big], \nonumber\\
{\cal E}_a^W &=&-\kappa_2\Big[  \frac{1}{2}\nabla_a K_\tau^2 
- K_\tau ( \eta_\perp^d \eta^c \nabla_a K_{dc} - K_{ad}\eta^d \nabla_c \eta_\perp^c \nonumber\\
&-& \eta^c K_a{}^b\nabla_b \eta_{\perp c} -  K_{ac} \eta^b_\perp \nabla_b \eta^c) \Big].
\end{eqnarray}
The corresponding Noether charge is identified as
\begin{eqnarray}
{\cal Q}^a_W &=&\kappa_2 \Big[  \frac{1}{2} K_\tau^2 \Phi^a  -\Phi K_\tau \eta^c \nabla^a \eta_{\perp c} 
-  K_\tau \eta^c \eta^a_\perp \nabla_c \Phi  \nonumber\\
&+&  \Phi \eta^a \eta^c_\perp \nabla_c K_\tau  + 2 \Phi K_\tau \eta^a \nabla_c \eta^c_\perp  \Big] .
\end{eqnarray}
In this case, the tangential deformation manifests itself only throughout the  twist energy density $K_\tau^2.$
\subsection{Stress and Torque tensor}
Under an infinitesimal translation, the twist stress tensor is obtained as
\begin{eqnarray}
{\bf f}^a_W&=& \kappa_2\Big[  K_\tau \left( \eta^c \eta^a_\perp K_c{}^b - \frac{1}{2} K_\tau g^{ab}   \right) {\bf e}_ b 
+( K_\tau \eta^c \nabla^a\eta_{\perp c} - \eta^a \eta^c_\perp \nabla_c K_\tau \nonumber\\
&-& 2K_\tau \eta^a \nabla_c \eta^c_\perp ) {\bf n} \Big].
\end{eqnarray} 
Notice that the tangential components $f^{ab}$, they do satisfy the condition $g_{ab}f^{ab}=0$, appeared
as a consequence
of the invariance under deformations $\delta {\bf X}= \lambda {\bf X}$.
In order to obtain the twist torque tensor we also need the derivative
\begin{eqnarray}
\nabla_a\Phi &=& \boldsymbol{\omega} \cdot ( {\bf e}_a \times {\bf n} + {\bf X}\times \nabla_ a {\bf n} ), \nonumber\\
&=&\boldsymbol{\omega} \cdot ( \varepsilon^b{}_a {\bf e}_ b + K_b{}^d {\bf X}\times {\bf e}_d ). 
\end{eqnarray}
In this way, the twist torque tensor is identified as
\begin{equation}
{\bf M}^a_W= {\bf X}\times {\bf f}^a_W + {\bf m}^a_W, 
\end{equation}
where ${\bf m}^a_W= \kappa_2 K_\tau \eta^a_\perp \ets_\perp$, is the intrinsic torque.\\
\subsection{Boundary conditions}
In order to obtain the boundary conditions we need to write
${\cal Q}^a_W l_a$, in terms of the independent deformations. On the boundary, the 
derivative of the normal deformation can be written as a linear combination 
of the independent derivatives
\begin{equation}
\nabla_a\Phi= l_a \nabla_l \Phi + T_a \dot\Phi,
\end{equation}
where  $\nabla_l= l^a\nabla_a \Phi$ and $\dot\Phi=T^a\nabla_a \Phi$.\\
Consequently, we get
\begin{eqnarray}
{\cal Q}^a_W l_a &=&\frac{\kappa_2}{2} K_\tau^2 \Psi_l +  \kappa_2 \Big[ -\Phi K_\tau \eta^c \nabla_l \eta_{\perp c} 
-  K_\tau \eta^c l_a\eta^a_\perp \nabla_c \Phi  \nonumber\\
&+&  \Phi l_a\eta^a \eta^c_\perp \nabla_c K_\tau  + 2 \Phi K_\tau l_a\eta^a \nabla_c \eta^c_\perp \Big].
\end{eqnarray}

Finally, the boundary conditions are  given by
\begin{eqnarray}
\sigma_b \kappa_g - \sigma &=& \frac{\kappa_2}{2}K_\tau^2, \nonumber\\
\sigma_b\kappa_n&=&
\kappa_2 \Big[ - K_\tau \eta^c \nabla_l \eta_{\perp c}  
+ l_a\eta^a \eta^c_\perp \nabla_c K_\tau  \nonumber\\
&+& 2 K_\tau l_a\eta^a \nabla_c \eta^c_\perp  + \frac{d}{ds}(\eta^c T_c l_a \eta^a_\perp)  \Big]\, \nonumber\\
0&=&  - \kappa_2 K_\tau \eta^c l_c l_a \eta^a_\perp. 
\end{eqnarray}
Notice that in this case there is no boundary condition arising from the boundary deformation along
the tangential direction.

\section{Bend energy}\label{BEND}
For evaluating  this component, it is convenient to define the vector field
$ {\bf B}= (\ets\cdot \nabla)\ets = {\cal A}^a {\bf e}_a - K_\eta {\bf n}, $
where we have written ${\cal A}^a= \eta^b\nabla_b \eta^a$, and $K_\eta= K_{ab}\eta^a\eta^b$.
Using this notation the bend energy reads more compact as
\begin{equation}
F_B= \frac{\kappa_3}{2} \int dA \, B^2,\label{BN}
\end{equation}
where $B^2= {\bf B}\cdot {\bf B}= {\cal A}^a{\cal A}_a+ K_\eta^2$.
The  deformation of the bend functional Eq. (\eqref{BN}), is given by (see  \ref{SSBB}):
\begin{equation}
\delta F_B = \int dA \, ({\cal E}^B \Phi +  {\cal E}^B_a\Phi^a) + \int dA\, \nabla_a{\cal Q}^a_B,
\end{equation}
where the Euler-Lagrange derivatives are given by
\begin{eqnarray}
\frac{{\cal E}^B}{\kappa_3} &=&- {\cal A}_b K^a{}_c\eta^c \nabla_a\eta^b - 
\nabla_c( K_a{}^b \eta^a\eta^c {\cal A}_b ) 
+ \nabla^a( K_\eta {\cal A}_a ) - K K_\eta^2 + K_\eta {\cal R}_G   \nonumber\\
&-& \nabla_a\nabla_b( \eta^a\eta^b K_\eta)+\frac{1}{2} K ( {\cal A}^a{\cal A}_a + K_\eta^2 ).\nonumber\\
\frac{{\cal E}^B_a}{\kappa_3} &=& {\cal R}_G {\cal A}_a - {\cal R}_G {\cal A}_c \eta^c \eta_a 
+ \nabla_c ( \eta^c {\cal A}_b \nabla_a \eta^b )
+ K_\eta \eta^c \eta^b \nabla_a K_{ cb}.
\end{eqnarray}
In this case, the Noether charge is
\begin{eqnarray}
{\cal Q}_B^a &=&\kappa_3\Big[ \frac{1}{2} ({\cal A}^b{\cal A}_b +K_\eta^2 ) \Phi^a 
-\eta^a ( \Phi^c {\cal A}_b \nabla_c \eta^b + K_\eta \eta^b \nabla_b \Phi )\nonumber\\
&+& \Phi \eta^a( 3 K^b{}_c {\cal A}_b \eta^c + K_\eta \nabla_b\eta^b 
+ \eta^b\eta^c\eta^d \nabla_b K_{cd}) \Big].  
\end{eqnarray}
\subsection{Stress and  Torque tensor}
Under an infinitesimal translation we get the bend stress tensor:
\begin{eqnarray}
{\bf f}^a_B&=& \kappa_3\Big[ \eta^a (  {\cal A}_b \nabla^c\eta^b + K_\eta \eta^b K_b{}^c  ) 
  - \frac{g^{ac}}{2}( {\cal A}^b{\cal A}_b  + K_\eta^2 ) \Big] {\bf e}_c \nonumber\\
&-& \kappa_3 \eta^a \Big[ 3 K_c{}^b \eta^c {\cal A}_b + K_\eta \nabla_b\eta^b 
+\eta^b\eta^c\eta^d \nabla_b K_{cd}\Big] {\bf n}.
\end{eqnarray}
As in the case of the splay and the twist stress tensor, the tangential 
components $f^{ab}$ do satisfy $g_{ab}f^{ab}=0$, as a consequence 
of invariance under deformations  $\delta{\bf X}= \lambda {\bf X}$.
Under an infinitesimal rotation we obtain the bend torque tensor as
\begin{equation}
{\bf M}^a_B= {\bf X}\times {\bf f}^a_B + {\bf m}^a_B,
\end{equation}
where ${\bf m}^a_B=\kappa_3 K_\eta \eta^a \ets_\perp $, is the 
corresponding intrinsic torque.
\subsection{Boundary conditions}
On the boundary, the projection of the Noether charge is given by 
\begin{eqnarray}
{\cal Q}^a_B l_a&=& \kappa_3 \left[  \frac{1}{2} ({\cal A}^b{\cal A}_b + K_\eta^2 )  - 
\eta^a l_a {\cal A}_b  \nabla_l \eta^b \right] \Psi_l 
-\kappa_3\,  \eta^a l_a {\cal A}_b T^c \nabla_c \eta^b\,  \Psi_T \nonumber\\
&+& \kappa_3[ 3 K_c{}^b \eta^c {\cal A}_b + K_\eta \nabla_b\eta^b 
+\eta^b\eta^c\eta^d \nabla_b K_{cd}    ]\Phi \nonumber\\
&-& \kappa_3 \eta^a l_a K_\eta \eta^b (  l_b \nabla_l \Phi + T_b \dot\Phi). \label{REF}
\end{eqnarray}
The boundary conditions are therefore 
\begin{eqnarray}
\sigma_b \kappa_g - \sigma &=&  \kappa_3 \left[  \frac{1}{2} ({\cal A}^b{\cal A}_b + K_\eta^2 )  - 
\eta^a l_a {\cal A}_b  \nabla_l \eta^b \right]  , \nonumber\\
\sigma_b \kappa_n &=& \kappa_3[ 3 K_c{}^b \eta^c {\cal A}_b + K_\eta \nabla_b\eta^b 
+\eta^b\eta^c\eta^d \nabla_b K_{cd}    \nonumber\\
&&\frac{d}{ds}(\eta^a l_a K_\eta \eta^b T_b)  ]\nonumber\\
0&=-&\kappa_3\,  \eta^a l_a {\cal A}_b T^c \nabla_c \eta^b \nonumber\\
0&=&-\kappa_3  K_\eta (\eta^b   l_b)^2.
\end{eqnarray}
The third boundary condition arises from the coefficient of the tangential deformation ($\Psi_T$ in Eq. (\ref{REF})),  being
valid even when the Canham-Helfrich elastic energy is taken into account (see \ref{HEL}).
One way to satisfy this condition consists to maintain the nematic field constant along the boundary ($\dot\eta^a=0$)
as in the splay case.

The above covariant theory is general for nematic membranes of arbitrary geometry considering liquid-crystal rigidities with non-vanishing values with respect to the bending stiffness of the flexible membrane ($\kappa_1, \kappa_2, \kappa_3 >> \kappa$ ). Using the tools of differential geometry to build upon the concepts of continuous mechanics, we have expanded the classical theory of curvature-elasticity of flexible membranes to account for internal ordering.   
In order to completeness, explicit account of the CH-contribution is required to describe the mechanical interplay between membrane ordering and flexibility (see \ref{HEL}). Specifically, the total current that generates the membrane on its boundary is defined as an invariant addition of conserved Noether charges 
($Q^a = Q^a_S + Q^a_W+Q^a_B+Q^a_{CH}$), which are associated to the corresponding tensors for membrane stress ${\bf f}^a = {\bf f}^a_S + {\bf f}^a_W+{\bf f}^a_B+{\bf f}^a_{CH}$  and torque $ {\bf M}^a = {\bf M}^a_S + {\bf M}^a_W+{\bf M}^a_B+{\bf M}^a_{CH}$ \cite{monroy}.  
Only in the limiting case of vanishing nematic ordering ($\kappa_1, \kappa_2, \kappa_3 << \kappa$), the Canhham-Helfrich theory is 
recovered \cite{guvenbook, stress, deserno}. In order to test for the practical applicability of our theory with specific membrane shapes, the embedding function must be subjected to adequate 2D-parametrization that allows explicit calculations of the strain field variables.

\section{Axial symmetry: Shape equation of revolution vesicles with nematic ordering}\label{AXIAL}
\begin{figure}[th]
\centering 
\includegraphics[width=3.4in]{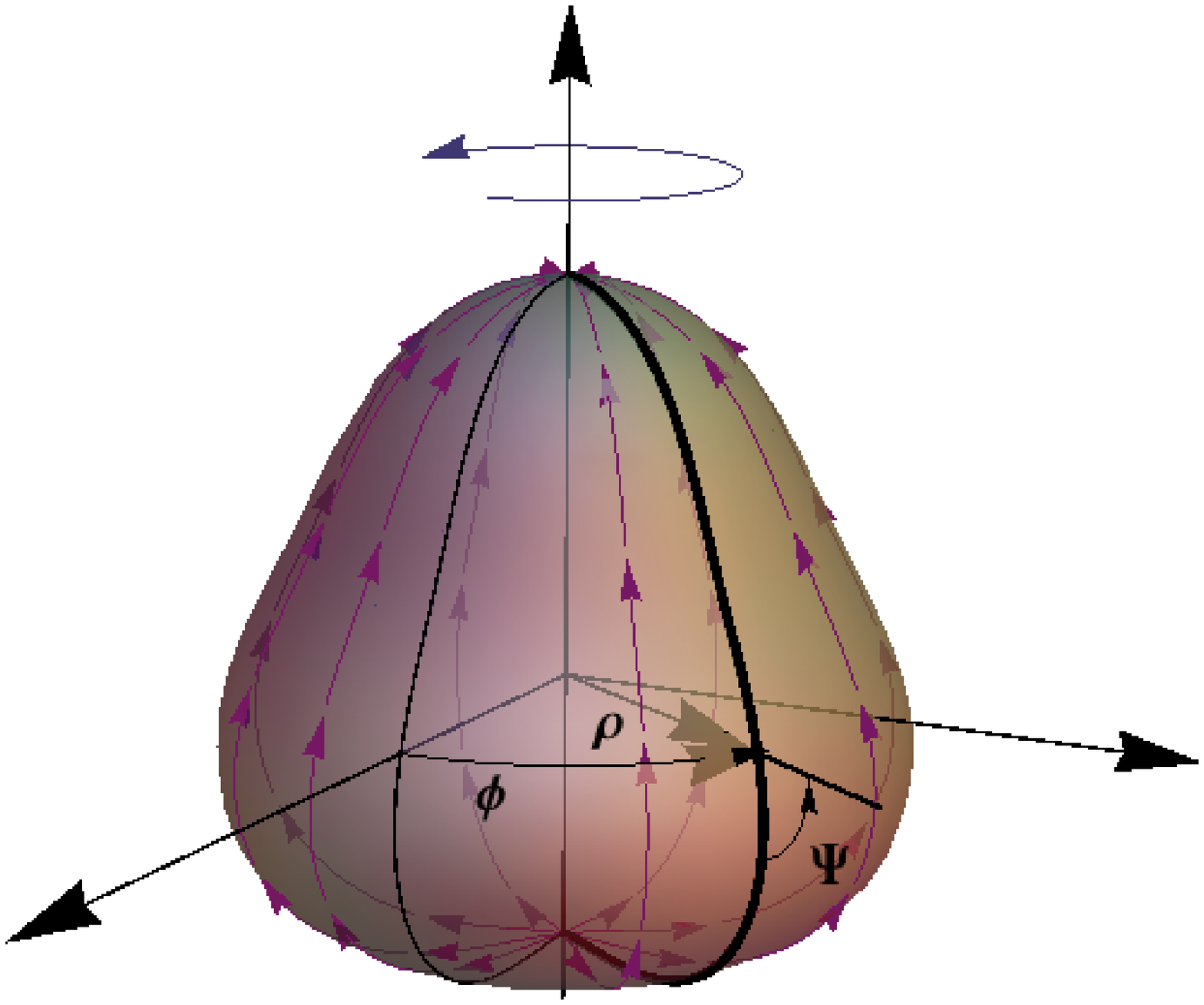}
\vspace{-2mm}
\caption{ A generic axially symmetric vesicle coated with a  nematic texture along the meridians. 
The angle $\phi\in [0, 2\pi]$ and $\rho\ge 0$.
The nematic texture includes a $+1$ topological defect at each pole. }
\vspace{-2mm}
\label{GG1}
\end{figure}

As an interesting example to test our theory, we address the particularly case of revolution surfaces with an uniaxial symmetry. 
Let the membrane be parametrized as
\begin{equation}
{\bf X}(l, \phi ) = ( \rho(l) \cos\phi, \rho(l)\sin\phi, z(l) ), 
\end{equation} 
where $l$ is the arc length along the meridians and $\phi\in[0, 2\pi]$, are presented in Figure~\ref{GG1}. 
For this parametrization, the surface
tangent vectors are given by
\begin{eqnarray}
{\bf e}_l &=& (\rho' \cos\phi, \rho'\sin\phi, z' )\nonumber\\
{\bf e}_ \phi&=& ( -\rho \sin\phi, \rho\cos\phi, 0 ),
\end{eqnarray}
where the symbol $'$ means derivative respect to $l$. Consequently, 
the induced metric is
\begin{equation}
g_{ab}d\xi^ad\xi^b= dl^2 + \rho^2 d\phi^2,
\end{equation}
where we have taken into account that $\rho'^2+ z'^2=1$. In this particular case, 
the only nontrivial Christoffel symbols are $\Gamma_{\phi l}^\phi= \rho'/\rho,$ and $\Gamma^l_{\phi\phi}=-\rho\rho' .$
The unit normal  to the surface, directed along $ {\bf e}_\phi\times {\bf e}_l$, is given by
\begin{equation}
{\bf n}=(z' \cos\phi, z'\sin\phi, -\rho' ).
\end{equation}
The second fundamental form can be written as
\begin{equation}
K_{ab}d\xi^ad\xi^b=-\frac{\rho''}{z'}dl^2 + \rho z' d\phi^2.
\end{equation}
For the trace of the extrinsic curvature we also have
\begin{equation}
K=g^{ab}K_{ab}= \frac{z'}{\rho} - \frac{\rho''}{z'}.
\end{equation}
Let us take a horizontal loop on the surface, its unit tangent ${\bf T}=(-\sin\phi, \cos\phi, 0 ) ={\bf e}_\phi/\rho $ and 
${\bf l}= {\bf T}\times {\bf n}= -{\bf e}_l $
complement the Darboux basis, such that: $T_l=0=T^l$, $T_\phi= \rho, T^\phi= 1/\rho$,   $l^l=-1=l_l$ 
and $l^\phi=0=l_\phi.$
Let a nematic texture oriented along the meridians such that
\begin{equation}
\ets =-( \rho' \cos\phi, \rho'\sin\phi, z' )=-{\bf e}_l.
\end{equation}
Its components are then $\eta^l=-1$ and $\eta^\phi=0$.
We also have $\ets_\perp= \ets \times {\bf n}=- {\bf T}$, so that $\eta_\perp^\phi= -1/\rho$, $\eta_\perp^l=0$.
Consequently, the divergence can be 
expressed as
\begin{equation}
\nabla_a \eta^a= -\frac{\rho'}{\rho},
\end{equation}
and $K_{ab}\eta^a l^b=-\rho''/ z' $.
We can write $l_a \eta^a= 1$ and we also see that 
$l^a\nabla_a \eta^b=0$, $T^a\nabla_a \eta^b=0$, and $\eta^a\nabla_a\eta^b=0$.

In this way, the splay force (per unit length) calculated  on a horizontal loop is given by 
${\bf f}^a_Sl_a= F^S_T{\bf T} + F^S_l {\bf l} + F^S_n{\bf n}$, where
the splay force projections are given by
\begin{eqnarray}
F^S_T&=& 0, \nonumber\\
F_l^S&=& -\frac{\kappa_1}{2} (\nabla_a\eta^a)^2, \nonumber\\
&=& -\frac{\kappa_1}{2} \left(\frac{\rho'}{\rho} \right)^2, \nonumber\\
F_n^S&=&\kappa_1\frac{\rho'}{\rho}\left( \frac{z'}{\rho} - \frac{2\rho''}{z'}   \right). 
\end{eqnarray}
The twist force 
$K_\tau= K_{ab}\eta^a \eta^b_\perp=0$, which implies that the twist tensor vanishes identically, ${\bf f}^a_W=0.$

To obtain the bend stress tensor, we find $K_\eta= K_{ab}\eta^a\eta^b=K_{ll}\eta^l\eta^l=-\rho''/ z'$.
We also have $K_{ab}T^a\eta^b=0$, and
$
\eta^b\eta^c\eta^d \nabla_b K_{cd}= -\nabla_l K_{ll}
=\left( \frac{\rho'' }{z'}  \right)'.
$
With these results at hand, we obtain the bend force on horizontal loops as ${\bf f}^a_B l_a = F^B_T{\bf T}+ F^B_l {\bf l} + F^B_n{\bf n}$,  
where the projections are given by
\begin{eqnarray}
F_T^B&=&0, \nonumber\\
F_l^B&=&  \frac{\kappa_3}{2}\left(\frac{\rho''}{z'} \right)^2,\nonumber\\
F_n^B&=&-\kappa_3 \left[\frac{\rho'}{\rho} \frac{\rho''}{z'}  + \left( \frac{\rho'' }{z'}  \right)'   \right]. 
\end{eqnarray}
Because the revolution symmetry, these expressions can be reduced to the shape of the generating curve. 
In terms of the inclination angle of the curve $\Psi(l)$ (see Figure \ref{GG1}), we can write:
\begin{eqnarray}
\rho'(l)&=& \cos\Psi, \nonumber\\
z'(l)&=& - \sin\Psi,
\end{eqnarray}
so, consequently: 
\begin{eqnarray}
F_l^S&=&-\frac{\kappa_1}{2} \left(\frac{\cos\Psi}{\rho}  \right)^2, \nonumber\\
F_n^S&=&-\kappa_1 \frac{\cos\Psi }{\rho}\left( \frac{\sin\Psi}{\rho} + 2\Psi'\right), \nonumber\\
F_l^W&=&F_n^W=0, \nonumber\\
F_l^B&=& \frac{\kappa_3}{2}\Psi'^2, \nonumber\\
F_n^B&=&-\kappa_3 \left( \frac{\cos\Psi}{\rho}\Psi' + \Psi'^2     \right).   \label{POP}
\end{eqnarray}
We recall that the corresponding Canham-Helfrich results
are given by \cite{deserno}
\begin{eqnarray}
F_l^{\rm CH}&=& -\sigma +\frac{\kappa}{2} \left[ \Psi'^2 -\left( \frac{\sin\Psi}{\rho}  \right)^2   \right], \nonumber\\
F_n^{\rm CH}&=& \kappa \left( \Psi'+ \frac{\sin\Psi}{\rho}  \right)', \label{OPQ}
\end{eqnarray}
which complement the total force field of the flexible membrane upon nematic ordering ($F = F^S + F^W + F^B + F^{CH}$). 
Particularly, If we take $\rho(0)=0$ (e.g. at the north pole), the equilibrium equation for the  revolution  shape  of  a closed vesicle 
with a meridian-oriented nematic texture (longitudinal director along the rotational symmetry axis) can be written as:
\begin{equation}
2\rho ( z' F_l - \rho' F_n  )= - P\rho^2, \label{MOL}
\end{equation}
where $P$ is the pressure difference between the outer medium and the vesicle interior.
Finally, taking into account summative contributions from Frank and Canham-Helfrich terms, using Eqs.
(\eqref{POP}) and (\eqref{OPQ}), Eq. (\eqref{MOL}) can be rewritten as:
 \begin{eqnarray}
&&\kappa (\cos\Psi) \Psi'' + \Big( \kappa_3 \cos\Psi -\frac{\kappa_3}{2}\sin\Psi -\frac{\kappa}{2}\sin\Psi \Big) \Psi^{'2}\nonumber\\
&& + ( 2\kappa_1 +\kappa_3 -\kappa )\frac{\cos^2\Psi}{\rho} \Psi' + \Big( \frac{3}{2}\kappa_1 \cos^2\Psi \nonumber\\
&&+\frac{\kappa}{2}\sin^2\Psi +\kappa \cos^2\Psi \Big)\frac{\sin\Psi}{\rho^2} + \sigma \sin\Psi  = -\frac{P}{2}\rho.\label{OOP}
\end{eqnarray}
The possible solutions for this equilibrium shape equation can be  obtained either numerically or by 
analytical procedures. This requires more specific work that 
is out of the scope of this paper. In quantitative terms, Eq. (\eqref{OOP}) tell us that the shape of the axisymmetric 
nematic vesicle with the meridian texture does not depend on the twisting rigidity, a fact already pointed out by 
Chen and Kamien  \cite{chen-kamien}.

\section{Discussion}\label{DIS}
Motivated by the physical understanding of internally-structured membranes, we have developed a generalized elasticity theory of a 2D-nematic governed 
by ordering-like distortion interactions. The membrane-embedded vectorial field of molecular directors is forced to lie in-plane (tangent) with a unitary modulus (inextensible), two essential features that constitutively define the 2D-nematic. The internal membrane organization is constructed as an in-plane vectorial ordering field, which is geometrically connected with the underlying strain field of curvature deformations. This nematic field is described by the surface-adapted Frank energy of liquid crystals \cite{monroy}, defined in terms of material rigidities that hinder the possible changes in the orientation of the molecular directors \cite{gennes}. The 2D-Frank energy is considered at zero-temperature thus do not including possible disordering effects arising from thermal fluctuations. The curvature deformations, and their intrinsically coupled variations in the nematic field, are described as infinitesimal changes of the surface embedding defined in terms of a parent strain field 
$\delta {\bf X}$. Such a genuine-elasticity concept blueprints the covariant theory here developed, which resolves the problem in a radically different pathway than the geometrical approach involved in our previous work \cite{monroy}. The present mechanical theory provides a complete, fully covariant, analytic framework for the systematic calculation of the membrane forces imposed by the structured nematic on the embedding surface. 

In our former geometric theory \cite{monroy}, we introduced ad hoc the tradeoff between molecular director and curvature as constraints in the orientation and the strength of the molecular director. That coupling was impinged to the surface geometry a posteriori, through Lagrange multipliers that compel the nematic directors to remain tangential in-plane with a unitary length. Such a geometrically-conditioned energy was then settled in a variational schema from which we obtained the equilibrium distribution of stresses compatible with the imposed geometrical constraints \cite{monroy}. Our current theory, however, is mechanically ab initio as builds upon an intrinsic strain field that, with the required properties, is embedded a priori in the surface geometry. Now, the surface nature of the nematic director is impregnated in the geometric structure of the deformation field (see Eq. (\eqref{WWE})), which injects the essential director characteristics to the embedding membrane; these are chosen to be tangent to the 
surface ($\ets\cdot {\bf n} = 0$) and to have a unitary length ($\ets\cdot \ets = 1$). This natural embedding in the membrane skeleton establishes a radical difference with respect to the geometric imposture previously considered \cite{monroy}, where the geometric characteristics of the nematic director were introduced in the Frank functional as auxiliary variables enforced to be conserved as external constraints. Whereas the former approach could appear a bit artificial, the current field theory is completely natural, as intrinsically embeds the structural characteristics in the field geometry. However, despite their evident operational differences, the two theories are completely equivalent as far the same structural ingredients result into identical expressions for the membrane forces.

From the current theory, we learn how surface geometry imposes changes in the director field, and viceversa, through affine connections directly related to the extrinsic curvature. The forces due to intrinsic deformations have been evaluated in terms of strains induced by the extrinsic membrane curvature on the surface metrics. Once all the possible modes of deformation have been included within a covariant setting, we have been able to construct the analytic edifice necessary to calculate the virtual works appeared upon splay, twist and bend of the director field. Building upon the geometry formalism previously defined \cite{stress, santiagog}, our mechanical theory extends far beyond the classical theory represented by the Canham-Helfrich model \cite{canham, helfrich}. With respect to the CH-theory developed for fluid 
membranes \cite{guvenbook, stress, deserno} (see \ref{HEL}), once coupled to the elastic energy of the flexible membrane characterized by the bending 
rigidity $\kappa$ \cite{monroy}, the current results represent an upgrade that accounts for internal nematic ordering; consequently, for $\kappa_i^{Frank} << \kappa$, our theory reduces indeed to the bare CH-theory \cite{monroy}. Specifically, to describe mechanics within the Frank elasticity setting (nematic-like), we have obtained the respective Euler-Lagrange (EL) derivatives and identified the Noether charges that represent the corresponding differentiable symmetries for every deformation mode that generates conserved currents at the membrane boundaries; these are (looking at Fig. \ref{FF1}), the rigid translation that generates the membrane stress and the rotation that generates the torque. From the material properties of the curvature-nematic field, the stresses and torques that are involved in the mechanical equilibrium have been described in compatibility with the variational principle. Further, the boundary conditions appear as a natural consequence of mechanical equilibrium at the membrane edges. With the results obtained from our nematic-embedding method, the mechanical route has been settled as generalized variations of the different components of the Frank energy describing the linear elastic response of the nematic texture within the three different distortion modes of the nematic director, these are, splaying, twisting and bending. Because analytic expressions for the EL-derivatives and Noether charges have been made available on these elasticity settings, we have been able to reveal how the significant changes in molecular director appear naturally coupled with the curvatures of the membrane. This analytic information paves the way not only for a quantitative determination of the equilibrium distribution of the membrane forces (stress and torque tensors) but also for a rigorous settlement of the membrane dynamics. Such a powerfulness is illustrated by the resolved example in Section \ref{AXIAL} dealing with the axial symmetry. In that specific case, a well-defined parametrization in spherical coordinates has provided simple analytic formulas for the equilibrium shape and the spatial distribution of the membrane forces in revolution cells. Our predictions are equivalent to previous results with minimal Delaunay surfaces \cite{chen-kamien}, but expand the analytic possibilities by merging membrane flexibility with nematic distortion in a curvature-ordering elasticity framework.

The theory of 2D-nematic elasticity developed in this work provides the analytic field framework required for studying the complex interplay (mechanics-geometry) between the equilibrium geometry and the internal ordering in nematic membranes at zero temperature. Because thermal fluctuations are not explicitly considered in our theory, the 2D-nematic is expected to retain orientational ordering at least in the topological sense of the Kosterlitz-Thouless theory at low temperature 
\cite{thouless2}, which grounds on a definition of order based in the overall topological properties of the system rather than on the classical definition of solid based on spatial correlations \cite{thouless1}. Ideally, we are dealing with a zero-temperature topologically ordered phase, a so-called liquid-crystalline texture, characterized by frozen spatial correlations that decrease with distance but retain some degree of nematic orientation. Even in infinite systems such a texture is assumed to be effectively frozen as far the orientational rigidities for splaying ($\kappa_1$), twisting ($\kappa_2$) and bending ($\kappa_3$) the nematic director are assumed to dominate upon the bending stiffness of the flexible membrane (kappa), which is usually higher than the thermal energy; in other words, the 2D-nematic is considered even more rigid than the rigid membrane where it is embedded, i.e. $\kappa_i \gtrsim \kappa >> k_BT$. Two additional factors are also crucial in determining 2D-nematic ordering in reduced geometries, these are membrane finiteness leading to constraining boundary conditions \cite{sandvik, izmail} and self-assembled topological defects in closed surfaces such as nematic shells \cite{napoli2012, napoli, ref1}.

On the one hand, the geometric constraints introduced by the membrane boundaries represent an important class of freezing interactions able to sustain orientational ordering in nematic membranes with a finite size. From the analysis of the boundary conditions in the considered nematic membranes, we deduced the presence of finite tension as a constraining condition that enforces the orientation of the equilibrium force field at the membrane edges. In general, the higher the membrane tension the higher the constraints against possible distortions on the nematic orientation. From the solutions obtained we deduce that the nematic field is highly constrained even at the free edge of a tensionless membrane ($\sigma = 0$). These boundary ordering effects are particularly relevant in small-systems where the finite size warrants remnant nematic ordering upon the orientational restrictions imposed by the membrane edges (see cartoon in Fig. \ref{FF1}). In larger systems at high temperature, even though the ordered nematic phase could be eventually destroyed by transverse fluctuations diverging with the system size \cite{mermin}, the 2D-nematic is expected to retain short-range order below some critical temperature corresponding to the Berezinski-Kosterlitz-Thouless (BKT) transition for unbinding topological 
defects \cite{thouless2}. Even at high temperature, Mermin argued that the thermal motion does not necessarily destroy orientational correlations at large distances, but even if long-range order disappears well above the BKT-transition, the direction taken by the nematic field in one region can be defined in terms of that taken in a previous region in the same neighborhood \cite{mermin}. Such a concept of short-range orientational ordering conveys in a local preservation of the liquid-crystalline order, which is the determining factor for the affine orientational interplay assumed in our theory between the nematic director field and the underlying curvature-field that imposes the geometric connection between them. 

On the other hand, the natural occurrence of topological defects in 2D-nematics represents an additional source of internal ordering leading to effective freezing of the orientational degrees of freedom \cite{nelsonbook}. In reduced geometries, particularly in nematic vesicles, the topological interactions between defects are known to impose strong entanglements among the different components of the nematic director field thus leading to highly-ordered textures \cite{napoli2010, ref1}. A paradigmatic example is the emergence of rigid meridian and parallel nematic field structures in spherical shells with two point-like $+1$ defects \cite{monroy}. Because the $+2$ Euler characteristic of the sphere, these two single defects repel each other and place equidistant in two opposite poles. The resulting “frozen” structures imposed by the mutually repulsing defects strictly determine the congruent topology of the two possible textures of the nematic field, either meridian orientation with the vectorial field aligned vertically north to south or parallel orientation with the vectorial field aligned horizontally along sphere parallels. Many other field-freezing interactions driven by topological defects are also possible in systems of variable size and geometry \cite{nelsonbook}. 

The two classes of orientational order, geometric or topological, respectively driven either by boundary conditions or by topological defects, may in general be referred to as freezing interactions able to sustain the orientational order necessary to define an effectively “frozen” vectorial structure in the 2D-nematic field. In the case of a nematic membrane, the disappearance of orientational order is associated with a transition from an elastic-like to a fluid-like mechanics. Even the appearance of BKT-defects as pairs of repulsive dislocations have finite distortion energy and must occur because of thermal excitation \cite{thouless2}. Only at sufficiently high temperatures, the largest pairs become unstable resulting in a fluid response to shear. A far the thermal energy overcomes the Frank energy in that case ($k_BT >> \kappa_i^{Frank}$), the anisotropic orientational ordering is completely destroyed thus resulting in a fluid-like behavior ultimately governed by the isotropic energy of the Canham-Helfrich functional (in case $\kappa >> \kappa_i^{Frank}$).
In this work, we build upon the well-known Canham-Helfrich model of fluid membranes that lead to the classical theory of isotropic curvature-elasticity 
\cite{guvenbook, deserno, fournier}. By expanding upon that mechanical theory for the disordered 2D-fluid, we have constituted the new theory of the 2D-nematic whose intrinsic anisotropy as a solid-like membrane with orientational order is supported on the assumed topological ordering considered at zero temperature. Although previous works have extensively attempted partial aspects of this problem  \cite{selinger, napoli2012, frank, santangelo, aharoni}, a closed theory had not been made available in a fully covariant form yet. Beyond the fundamental interest of our development, further analyses could be relevant to different fields of science and technology where the enhanced response of the hybrid membrane is encoded on its internally ordered structure; from the physics of soft materials based on liquid crystals \cite{ref1, ref2}, to the biophysics of cellular membranes \cite{ref5}, through the engineering of "smart" shells based on composite wired structures \cite{ref4}. By using a limited number of constitutive parameters, the new theory is predictive about the complex morphological remodeling occurred on internally structured surfaces upon generalized deformation. We have introduced the essential mechanics that captures the most generic effects of nematic-like ordering inside a flexible structure, thus providing an ample analytic platform to study a variety of membrane processes in terms of internal ordering.

\section{Conclusions}\label{CON}
This work addresses the formal theory of curvature-elasticity of nematic membranes. The novelty represented by the current work means an advance in the theoretical understanding of flexible membranes with nematic ordering. A promising analytic gate is opened towards the rational mechanics of internally ordered membranes. Further work on the predictions of this theory with specific examples is ongoing.

\section*{Acknowledgments} 
JAS thanks to Prof. Francisco Monroy  hospitality at Universidad Complutense de Madrid where this work was done.
The authors wish to thank the discussions and the interest  of C.M. Corona-Rubio.
The work was supported in part by CONACyT under Becas Sab\'aticas en el Extranjero (to JAS), and by Ministerio de Econom\'ia y Competitividad (MINECO, Spain) under grant FIS2015-70339-C2-1-R and by Comunidad de Madrid under grants S2013/MIT- 2807, P2018/NMT4389 and Y2018/BIO5207 (to FM).

\appendix

\section{Nematics on flat membrane}\label{NNPLANE}
Let us consider a region of the plane $XY$ such that the membrane shape ${\bf x}= {\bf X}(x, y)=(x, y, 0)$ and let
$\delta {\bf X}= ( \sin x \cosh y , \cos x \sinh y, 0  )$, to be an in-plane deformation of this plane. The unit tangent
vectors are given by ${\bf e}_x=(1, 0, 0)$ and ${\bf e}_y=(0, 1, 0)$, whereas the unit normal
${\bf n}=(0, 0, 1)$. According to Eq. (\eqref{DEFOR}), the tangential deformations are $\Phi^x= \sin x \cosh y $, and
$\Phi^y= \cos x \sinh y$, while the normal deformation vanishes,  $\Phi=0$. \\
If we start with a uniform nematic texture with director $\ets={\bf e}_y$ so that $\eta^y=1$ and $\eta^x=0$, 
we find its deformation  as 
\begin{equation}
\delta\eta^x= -\sin x \sinh y,  \,\,\,\,\,\,\, \,  \delta\eta^y= -\cos x \cosh y. \label{MNN}
\end{equation}
In this planar setting, the deformed nematic texture has a non-trivial splay energy $(\partial_a\eta^a)^2= 4( \cos^2 x) \sinh^2 y $.
For example,  in the case of  the nematic director oriented along the $x=\pi/2$-axis,
we found $\delta\eta^x=-\sinh y $ and $\delta\eta^y=0$ (with increasing length as a function of $y$).
In this way, the nematic director is transformed by changing its length and direction. 
However, at this point we recall that the tangent  vectors ${\bf e}_a=\partial_a {\bf X}$, do depend 
on the shape membrane, and they also undergo deformations  according to 
\begin{equation}
\delta{\bf e}_y=\sin x\sinh y\, {\bf e}_x + \cos x\cosh y \, {\bf e}_y. \label{GHU}
\end{equation}
Consequently, the entire deformation of the nematic director $\ets={\bf e}_y$, is given by
\begin{eqnarray}
\delta\ets &=& \delta\eta^a {\bf e}_a + \eta^a\delta{\bf e}_a, \nonumber\\
&=&\delta\eta^x {\bf e}_x + \delta\eta^y{\bf e}_y + \delta{\bf e}_y,
\end{eqnarray}
which vanishes when Eqs. (\eqref{MNN}) and (\eqref{GHU}) for in-plane deformations are substituted.

\section{Splay}\label{SSPP}
The obtain the deformation of  divergence of the nematic director, we write 
\begin{eqnarray}
\delta (\nabla\cdot \ets) &=&\delta g^{ab} ({\bf e}_a \cdot \nabla_b\ets  ) + g^{ab} \delta {\bf e}_a\cdot \nabla_a \ets\nonumber\\
&+& g^{ab}{\bf e}_a \cdot \delta \nabla_b \ets. \label{DERS}
\end{eqnarray}
The first term in Eq. (\eqref{DERS}) gets 
\begin{eqnarray}
\delta g^{ab} ({\bf e}_a\cdot \nabla_b\ets ) &=&-(2\Phi K^{ab} + \nabla^a\Phi^b + \nabla^b\Phi^a) \nabla_b\eta_a, \nonumber\\
&=&-2\Phi K_a{}^b \nabla_b \eta^a \nonumber\\
&-&\nabla_a ( \Phi^b \nabla_b\eta^a) + \Phi^b \nabla^a ( \nabla_b\eta_a  )\nonumber\\
&-&\nabla_b( \Phi^a \nabla_b\eta_a )+ \Phi^a \nabla^b ( \nabla_b \eta_a). \label{DS1} 
\end{eqnarray}
The second term can be developed as
\begin{eqnarray}
g^{ab}\delta{\bf e}_a\cdot \nabla_a\ets&=&\nabla^a\eta^c ( \nabla_a\Phi_c + \Phi K_{ac} )\nonumber\\
&-&\eta^c K^a{}_c (\nabla_a \Phi - K_{ad}\Phi^d ), \nonumber\\
&=&\Phi K^a{}_b \nabla_a \eta^b +\eta^c \Phi^d K^a{}_c K_{ad}  \nonumber\\
&+& \nabla_a( \Phi_c \nabla^a\eta^c  ) - \Phi^c \nabla_a ( \nabla^a\eta_c)  \nonumber\\
&-&\nabla_a( \Phi \eta^c K^a{}_c  ) + \Phi \nabla_a( \eta^c K^a{}_c ), \label{DS2}
\end{eqnarray}
and finally,  the  third term can be written as
\begin{eqnarray}
g^{ab}{\bf e}_a \cdot \delta \nabla_b \ets &=& (\nabla_c\Phi^b + \Phi K^b{}_c ) \nabla_b\eta^c \nonumber\\
&-& \eta^c K^a{}_c ( -\nabla_a \Phi + K_{ad}\Phi^d )  \nonumber\\
&+& \delta (\nabla_a\eta^a).\nonumber\\
&=& \Phi K_c{}^b \nabla_b\eta^c - \eta^c \Phi^d K^a{}_c K_{ad}\nonumber\\
&+& \nabla_c( \Phi^b \nabla_b \eta^c ) - \Phi^b \nabla_c (  \nabla_b \eta^c )\nonumber\\
&+& \nabla_a ( \Phi \eta^c K^a{}_c )- \Phi \nabla_a( \eta^c K^a{}_c  )\nonumber\\
&+& \delta (\nabla_a\eta^a). \label{DS3}
\end{eqnarray}
When the three terms, Eqs. (\eqref{DS1}), (\eqref{DS2}) and (\eqref{DS3}) are added together, they all cancel 
out except the deformation of the divergence $\delta(\nabla_a\eta^a)$.
Explicitly writing the covariant derivative $\nabla_a \eta^a$,  its deformation can be obtained as
\begin{eqnarray}
\delta \nabla_a \eta^b &=& \nabla_a \delta \eta^b +  \delta\Gamma_{ac}^b \eta^c,
\end{eqnarray}
and then we can write the commutator among the deformation operator and the covariant derivative, we have
\begin{equation}
[\delta, \nabla_a] \eta^b=  \delta\Gamma_{ac}^b \eta^c,\label{CMM}
\end{equation}
where on the right hand side we have deformations of the Christoffel symbols, them  can be calculated in 
terms of the induced metric deformations:
\begin{equation}
\delta\Gamma_{ab}^c= \frac{1}{2} g^{cd} (\nabla_b \delta g_{ad} + \nabla_a \delta g_{bd} - \nabla_d \delta g_{ab} ),
\end{equation}
where $\delta_\perp g_{ab}= 2K_{ab} \Phi $ and $\delta_\parallel g_{ab}= \nabla_a\Phi_b + \nabla_b\Phi_a,$
should be used. In this way,  the tangential deformation gets into
\begin{eqnarray}
\delta_\parallel\Gamma_{ab}^c 
&=& \nabla_a \nabla_b \Phi^c + {\cal R}_G ( g_{ab}\Phi^c -\delta^c_a \Phi_b ),
\end{eqnarray}
where ${\cal R}_G$ is the Gaussian curvature of the surface.
On the other hand we can obtain
\begin{eqnarray}
\nabla_a \delta\eta^b &=& -\nabla_a\eta^c ( \nabla_c \Phi^b + \Phi K_c{}^b) \nonumber\\
&-& \eta^c [ \nabla_a \nabla_c \Phi^b -\nabla_a( \Phi K_c{}^b)  ].
\end{eqnarray}
Hence, by using  the commutator Eq. (\eqref{CMM}), we can write the tangential deformation 
of the covariant derivative as
\begin{eqnarray}
\delta_\parallel \nabla_a\eta^b = - \nabla_a\eta^c ( \nabla_c \Phi^b) 
+ {\cal R}_G (\Phi^b\eta_a - \delta_a^b \Phi_c\eta^c ).\label{DA}
\end{eqnarray}
Similarly, the normal deformation can be obtained  as
\begin{equation}
\delta_\perp \nabla_a\eta^b= -\Phi K_c{}^b  \nabla_a \eta^c + K_a{}^b \eta^c \nabla_c \Phi 
- K_{ac}\eta^c \nabla^b\Phi. \label{DB}
\end{equation}

\section{Twist}\label{TTSS}
Deformation of the twist energy density is given by
\begin{eqnarray}
K_\tau \delta K_\tau &=&( \delta g^{ab}) K_\tau \ets \cdot ({\bf e}_a \times \nabla_b \ets)
+ g^{ab} K_\tau (\delta\ets) \cdot  ({\bf e}_a \times \nabla_b \ets)\nonumber\\
&+& g^{ab} K_\tau \ets \cdot  [ (\delta{\bf e}_a ) \times \nabla_b \ets ]
+ g^{ab} K_\tau \ets \cdot  ({\bf e}_a \times \delta \nabla_b \ets). \label{TWW}
\end{eqnarray}
The first line in 	Eq. (\eqref{TWW}) can be developed as
\begin{eqnarray}
\delta g^{ab} \ets \cdot ( {\bf e}_a \times \nabla_b \ets ) K_\tau  
 &=& -2\Phi K K_\tau^2 - \nabla_a( \eta^c K_\tau K_{bc} \eta^a_\perp \Phi^b)\nonumber\\ 
&&+ \nabla^a( \eta^c K_\tau K_{bc} \eta_{\perp a} ) \Phi^b 
+\nabla^b ( \eta^c K_\tau K_{bc}\eta_{\perp a}  )\Phi^a \nonumber\\
&&- \nabla_a ( \eta^c K_\tau K^a{}_c \eta_{\perp b} \Phi^b), 
\end{eqnarray}
where we have used the deformation of the inverse of the induced metric, 
$\delta g^{ab}= - (2\Phi K^{ab}+ \nabla^a\Phi^b + \nabla^b\Phi^a)$.
The second line in Eq. (\eqref{TWW}) is given by
\begin{eqnarray}
g^{ab}K_\tau \delta\ets\cdot ({\bf e}_a \times \nabla_b\ets)&=&
-K_\tau K_{dc} \eta^d \Phi^c \nabla_a\eta_\perp ^a 
- \nabla_d( K_\tau  \eta^d \nabla_b \eta^b_\perp) \Phi \nonumber\\
&+& \nabla_a ( K_\tau \eta^a \nabla_c\eta^c_\perp  \Phi).
\end{eqnarray}
The third line can be developed as
\begin{eqnarray}
g^{ab} K_\tau\ets\cdot ( \delta{\bf e}_a\times \nabla_b \ets)
&=& \Phi K K_\tau^2 + \Phi \nabla_a (K_\tau \eta^c \nabla^a\eta_{\perp c})\nonumber\\
&-& \Phi^d \nabla_a(K_\tau \eta^c K^a{}_c\eta_{\perp d} ) + \Phi^d \eta^c K_\tau K^a{}_d \nabla_a \eta_{\perp c}  \nonumber\\
&+&\nabla_a(\eta^c K_\tau K_c{}^a\eta_{\perp d} \Phi^d ) - \nabla_a (\Phi K_\tau \eta^c \nabla^a \eta_{\perp c}   ).
\end{eqnarray}
We have used  that
\begin{eqnarray}
\delta {\bf e}_a \times \nabla_b \ets &=& ( \nabla_a \Phi^d +\Phi K_a{}^d )   (\nabla_b \eta_{\perp d}) {\bf n}
-  ( \nabla_a \Phi^d +\Phi K_a{}^d ) \eta^c K_{bc} \varepsilon_{ed} {\bf e}^e\nonumber\\
&-& (\nabla_a \Phi - K_{ad}\Phi^d )\nabla_b\eta_{\perp e} {\bf e}^e,
\end{eqnarray}
so that we get
\begin{eqnarray}
\ets\cdot (\delta {\bf e}_a \times \nabla_b \ets) = ( \nabla_a \Phi^d +\Phi K_a{}^d ) \eta^c K_{bc} \eta_{\perp d} 
- (\nabla_a \Phi - K_{ad}\Phi^d )\eta^c \nabla_b \eta_{ \perp c}.
\end{eqnarray}
The fourth line in Eq. (\eqref{TWW})  involves (\eqref{DDNS}), so that
\begin{eqnarray}
K_\tau \ets \cdot ( {\bf e}_a \times \delta\nabla_b \ets ) &=& - \nabla_c ( \Phi K_\tau  \eta^a_\perp \nabla_a \eta^c ) 
+ \Phi \nabla_c ( K_\tau \eta^a_\perp \nabla_a\eta^c  )\nonumber\\
&+& \Phi^d \nabla_b\eta^c K_\tau K_{cd} \eta_{\perp a} 
+K_\tau \delta (\eta^c K_{bc} ) \eta_{\perp a},\label{GOP}
\end{eqnarray}
here we recall the deformations of the extrinsic curvature
\begin{eqnarray}
\delta_\parallel K_{ab}&=& \Phi^c\nabla_c K_{ab} + K_{ac}\nabla_b\Phi^c + K_{bc}\nabla_a \Phi^c    , \nonumber\\
\delta_\perp K_{ab}&=&- \nabla_a\nabla_b \Phi + ( KK_{ab}- {\cal R}_G g_{ab} ) \Phi. \label{DKP}
\end{eqnarray}
Thus, after some simplifications and  grouping some terms we have the last term in Eq. (\eqref{GOP})
\begin{eqnarray}
K_\tau \eta^a_\perp \delta ( \eta^c K_{ac} ) &=&  \Phi^d K_\tau \eta^a_\perp \eta^c \nabla_d K_{ac} + 
\nabla_a ( \Phi^d K_\tau K_{cd} \eta^a_\perp\eta^c  )\nonumber\\
&-& \Phi^d \nabla_a ( \eta^a_\perp \eta^c K_\tau K_{cd} ) - \nabla_a ( K_\tau \eta^c \eta^a_\perp \nabla_c\Phi )\nonumber\\
&+& \nabla_c[ \Phi \nabla_a( K_\tau \eta^c \eta^a_\perp)] - \Phi \nabla_c \nabla_a( K_\tau \eta^c \eta^a_\perp ).\nonumber
\end{eqnarray}
\section{Bend}\label{SSBB}
Deformation of the bend functional Eq. (\eqref{BN}), is given by
\begin{equation}
\delta F_B= \frac{\kappa_3}{2}\int \delta (dA) B^2 +\kappa_3 \int dA\, {\bf B}\cdot \delta {\bf B}. 
\end{equation}
Thus the bending energy deformation can be written in terms of 
\begin{eqnarray}
\delta {\bf B} = \delta{\cal A}^b {\bf e}_b + {\cal A}^b \delta {\bf e}_ b   
-\delta K_\eta {\bf n} - K_\eta \delta{\bf n}.
\end{eqnarray}
We have
\begin{eqnarray}
{\bf B}\cdot \delta {\bf B} &=& {\cal A}_b\delta {\cal A}^b 
+ {\cal A}^c {\cal A}^b {\bf e}_c\cdot \delta {\bf e}_ b  
+ K_\eta \delta K_\eta.\nonumber\\
&=&{\cal A}_a \delta {\cal A}^a + K_\eta \delta K_\eta  + \underbrace{\nabla_b ( \Phi_c {\cal A}^c {\cal A}^b )}_{1} 
-\underbrace{ \Phi_c \nabla_b( {\cal A}^c {\cal A}^b )}_{2} \nonumber\\
&+& \underbrace{\Phi K_{bc} {\cal A}^b {\cal A}^c}_{3},\label{DDJ}
\end{eqnarray}
where we have used that ${\bf e}_c\cdot\delta{\bf e}_b= \nabla_b\Phi_c + \Phi K_{bc}$.
By using Eqs. \eqref{DA} and \eqref{DB} we can write
\begin{eqnarray}
{\cal A}_ b\delta {\cal A}^b &=& - \Phi K^c{}_a \eta^c {\cal A}_b \nabla_a\eta^b + \Phi^b {\cal A}_b {\cal R}_G\nonumber\\
&-& \Phi^b \eta_b {\cal A}_c \eta^c {\cal R}_G -\underbrace{ \Phi K_{bc} {\cal A}^b{\cal A}^c}_{3}\nonumber\\
&-&\nabla_c(\Phi^a \eta^c {\cal A}_b\nabla_a \eta^b   )  + \Phi^a \nabla_a( \eta^c {\cal A}_b\nabla_a\eta^b )\nonumber\\
&-&\underbrace{\nabla_c( \Phi^b {\cal A}_b{\cal A}^c )}_{1} +
\underbrace{\Phi^b \nabla_c ( {\cal A}_b {\cal A}^c  )}_{2}\nonumber\\
&+& \nabla_c( \Phi K_a{}^b \eta^a{\cal A}_b \eta^c) - \Phi \nabla_c ( K_a{}^b \eta^a {\cal A}_b \eta^c  )\nonumber\\
&-& \nabla_b (\Phi K_\eta {\cal A}^b) + \Phi \nabla_b ( K_\eta {\cal A}^b ). \label{HJL}
\end{eqnarray}
Note that the terms marked $1, 2, 3$ cancel each other in Eqs. \eqref{DDJ} and \eqref{HJL}.
On the other hand, we see that
\begin{equation}
\delta K_\eta=\eta^a\eta^b\delta K_{ab}  + 2K_{ab}\eta^b \delta \eta^a, 
\end{equation}
and using the deformation of the extrinsic curvature Eq. (\eqref{DKP}), we can write the second term in Eq. (\eqref{DDJ}) as
\begin{eqnarray}
K_\eta \delta K_\eta &=& \Phi^c K_\eta \eta^a \eta^b \nabla_c K_{ab} -\Phi KK_\eta^2 + \Phi K_\eta {\cal R}_G\nonumber\\
&-& \Phi \nabla_b\nabla_a (K_\eta \eta^a\eta^b ) - \nabla_a( K_\eta \eta^a \eta^b \nabla_b\Phi )\nonumber\\
&+& \nabla_b[\Phi \nabla_a( K_\eta \eta^a \eta^b ) ].  
\end{eqnarray}

\section{Boundary conditions of the Canham-Helfrich model}\label{HEL}
For a closed vesicle, the integrated Gaussian curvature is a topological invariant determined
by the Gauss-Bonnet theorem \cite{docarmo}, however 
it has a finite contribution in the presence of boundaries. Let us consider us the Canham-Helfrich energy 
with a boundary with linear tension $\sigma_b$:
\begin{equation}
F_{CH}= \frac{\kappa}{2}\int dA(K-K_0 )^2 + \overline{\kappa} \int dA {\cal R}_G + \sigma \int dA + \sigma_b \oint ds,   \label{CEH}
\end{equation}
where  $ds$ is the  arc length element on the boundary curve.
Under $\delta{\bf X}= \Phi^a {\bf e}_a + \Phi {\bf n}$, 
the   deformation of the area element is given by
\begin{equation}
\delta dA=dA ( K\Phi +\nabla_a \Phi^a ),
\end{equation}
while the deformation of the curvature $K$ can be obtained by using Eqs. (\eqref{DKP}); for the total change,  we have \cite{santiago-defor}
\begin{equation}
\delta K= - \Delta\Phi + \Phi^a\nabla_a K+  (2 {\cal R}_G -K^2  )\Phi.
\end{equation}
Similarly, deformation of the Gaussian curvature is given by
\begin{equation}
\delta {\cal R}_G=\Phi^a \nabla_a {\cal R}_G - K {\cal R}_G \Phi + \nabla_a(h^{ab}\nabla_b \Phi),
\end{equation}
where $h^{ab}=K^{ab}-g^{ab} K$.
With these results at hand, after  some integration by parts, we obtain the deformation of the  Canham-Helfrich energy (\eqref{CEH}) as
\begin{equation}
\delta F_{CH}= \int dA \, {\cal E}_{CH} \Phi+ \int dA \nabla_a Q^a_{CH},\label{BOOD}
\end{equation}
where 
\begin{equation}
{\cal E}_{CH}= -\kappa \Delta K - \frac{\kappa}{2}(K-K_0)[ K(K+K_0)- 4{\cal R}_G ] +\sigma K,
\end{equation}
is the Euler-Lagrange derivative; we observe that the Gaussian rigidity does not appear. 
Because the last term in Eq. (\eqref{BOOD}) can be written as the boundary term
\begin{equation}
\int dA \, \nabla_a Q_{CH}^a=\int ds \, l_aQ_{CH}^a,
\end{equation}
where
\begin{eqnarray}
{\cal Q}^a_{CH} l_a &=& \kappa [ \Phi \nabla_l K -  (K-K_0)\nabla_l \Phi  +\frac{1}{2} (K-K_0)^2\Psi_l ]  \nonumber\\
&+& \overline\kappa [ {\cal R}_G \Psi_l - K_{ab}T^aT^b\nabla_l \Phi + K_{ab}l^aT^b \dot\Phi] + \sigma \Psi_l,
\end{eqnarray}
which is written in terms of the generalized deformation fields, $\Phi$ for the normal strain  and $\Psi_l$ for the in-plane longitudinal strain 
(Eq. \eqref{STRATAN}). 
Let us notice as the CH-Noether charge contains three contributions arising from the pure bending stress, the Gaussian stiffness and the membrane tension, respectively. They generate the conserved currents that define the mechanics of the inextensible but flexible membrane. 

Consequently, the boundary conditions for the CH-energy (\eqref{CEH}) are then given as~\cite{santiago-edge}
\begin{eqnarray}
\frac{\kappa}{2} (K-K_0 )^2 &=& -\overline\kappa\, {\cal R}_G +\sigma_b \kappa_g -\sigma,  \nonumber\\
\kappa (K-K_0) &=& \overline\kappa \kappa_n, \nonumber\\
\kappa\nabla_l K &=&\dot\tau_g + \sigma_b \kappa_n, \nonumber\\
\end{eqnarray}
where we have taken into account the definitions of the Darboux frame in Eq. (\eqref{DAR}). 
Despite the irrelevance of the Gaussian rigidity in the global shape of closed geometries at equilibrium, 
the above equations point out the non-trivial influence of $\overline\kappa$ in determining membrane mechanics on the boundaries.



\end{document}